\documentclass{article}

\usepackage{amsmath,amssymb,amsthm, graphicx,latexsym,epic}
\usepackage{cite,enumerate,float,indentfirst}
\usepackage{color,tikz}
\usetikzlibrary{arrows,snakes,backgrounds}

\def\be{\begin{eqnarray}}
\def\ee{\end{eqnarray}}
\def\nn{\nonumber}

\def\Gns{{\rm Gns}}

\newcommand{\beq}{\begin{equation}}
\newcommand{\eeq}{\end{equation}}
\newcommand{\beqa}{\begin{eqnarray}}
\newcommand{\eeqa}{\end{eqnarray}}
\newcommand{\CR}{\nonumber \\}
\newcommand{\gq}{\mathfrak{q}}

\newcommand{\floor}[1]{\lfloor#1\rfloor}
\newcommand{\lam}{\lambda}
\newcommand{\m}{\mu}

\definecolor{red}{rgb}{1,0,0}
\definecolor{orange}{rgb}{1,0.5,0}
\definecolor{violet}{rgb}{0.7,0,1}

\hoffset= - 1.0in         

\begin{document}

\title{\bf
Elliptic lift of the Shiraishi function
as a non-stationary
double-elliptic function
}

\author{
{\bf Hidetoshi Awata$^a$}\footnote{awata@math.nagoya-u.ac.jp},
\ {\bf Hiroaki Kanno$^{a,b}$}\footnote{kanno@math.nagoya-u.ac.jp},
\ {\bf Andrei Mironov$^{c,d,e}$}\footnote{mironov@lpi.ru; mironov@itep.ru},
\ and \  {\bf Alexei Morozov$^{f,d,e}$}\thanks{morozov@itep.ru}
\date{ }
}

\maketitle

\vspace{-6.0cm}

\begin{center}
\hfill FIAN/TD-01/20\\
\hfill IITP/TH-01/20\\
\hfill ITEP/TH-01/20\\
\hfill MIPT/TH-01/20
\end{center}

\vspace{4.7cm}

\begin{center}
$^a$ {\small {\it Graduate School of Mathematics, Nagoya University,
Nagoya, 464-8602, Japan}}\\
$^b$ {\small {\it KMI, Nagoya University,
Nagoya, 464-8602, Japan}}\\
$^c$ {\small {\it Lebedev Physics Institute, Moscow 119991, Russia}}\\
$^d$ {\small {\it ITEP, Moscow 117218, Russia}}\\
$^e$ {\small {\it Institute for Information Transmission Problems, Moscow 127994, Russia}}\\
$^f$ {\small {\it MIPT, Dolgoprudny, 141701, Russia}}
\end{center}

\vspace{.0cm}

\begin{abstract}
As a development of \cite{AKMMdell1}, we note that the ordinary
Shiraishi functions have an insufficient number of parameters
to describe generic eigenfunctions of double elliptic system (Dell).
The lacking parameter can be provided by substituting elliptic
instead of the ordinary Gamma functions in the coefficients
of the series. These new functions (ELS-functions) are conjectured to be functions governed by compactified DIM networks
which can simultaneously play the three roles:
solutions to non-stationary Dell equations,
Dell conformal blocks with the degenerate field (surface operator) insertion,
and the corresponding instanton sums in $6d$ SUSY gauge theories with adjoint matter.
We describe the basics of the corresponding construction and
make further conjectures about the various limits and dualities
which need to be checked to make a precise statement about
the Dell description by double-periodic network models with
DIM symmetry. We also demonstrate that the ELS-functions  provide symmetric polynomials,
which are an elliptic generalization of Macdonald ones, and
compute the generation function of the elliptic genera of the affine
Laumon spaces. In the particular $U(1)$ case, we find an explicit
plethystic formula for the $6d$ partition function, which is a
non-trivial elliptic  generalization of the $(q,t)$ Nekrasov-Okounkov
formula from $5d$.
\end{abstract}

\section{Introduction}

Nekrasov's extension \cite{Nek1,Nek2} of Seiberg-Witten theory \cite{SW1,SW2}
describes instanton sums
(LMNS integrals \cite{LMNS1,LMNS2})
in the $\Omega$-background deformed supersymmetric Yang-Mills theories.
Like Seiberg-Witten theory itself \cite{GKMMM,DW}, this extension can be encoded in terms
of the theory of integrable systems: in particular, in the Nekrasov-Shatashvili limit \cite{NS},
when one of the two $\Omega$-deformation $\epsilon$-parameters vanishes,
an AGT related conformal block with a degenerate field (surface operator)
insertion \cite{TM,MMMsurfop} satisfies the Baxter equation \cite{MM1,MM2,MMS},
i.e. the stationary Schr\"odinger equation for the quantized integrable systems in separated variables.
The full-fledged $\Omega$-deformation is supposed to be related to
non-stationary Schr\"odinger equation \cite{Braver, BE,MMMsurfop}.
An adequate language to describe the entire set of problems,
including the AGT relations \cite{AGT1,AGT2,AGT3} and extended Virasoro symmetries \cite{CFT},
is that of network models \cite{network1,network2}, which is basically a representation theory of the
Ding-Iohara-Miki (DIM) algebras \cite{DIM1,DIM2}, a far-going unification of the Virasoro and $W$ algebras.
In this SYM/integrability dictionary, the most rich and interesting
is the still mysterious double-elliptic (Dell) system \cite{BMMM,Dell,MMd}.
At the level of integrable systems of the Calogero-Ruijsenaars type, it possesses
double periodicity (ellipticity) in both momenta and coordinates \cite{BMMM,Dell}.
In Seiberg-Witten theory, it corresponds to the $6d$ SYM system
with adjoint matter hypermultiplet \cite{MMd,Gleb}
(adjoint is known to gives rise to the elliptic Calogero-Ruijsenaars systems \cite{DW,IM,BMMM5},
while the other torus comes from the compactification from $6d$ to $4d$).
In the language of network systems, one needs to compactify the network
in both the vertical and horizontal directions \cite{Mironov:2015thk,Mironov:2016cyq,AKMMSZ}, see Fig.1.

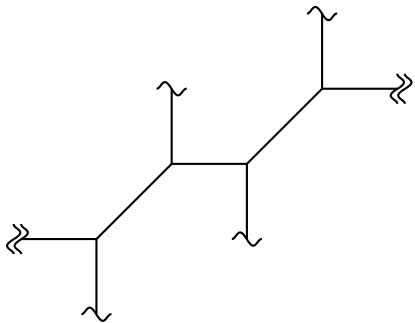
\begin{figure}
\begin{center}
\begin{tikzpicture}[thick]
\draw[black] (0,0) -- (1,0);
\draw[black] (1,0) -- (1,-1);
\draw[black] (1,0) -- (2,1);
\draw[black] (2,1) -- (2,2);
\draw[black] (2,1) -- (3,1);
\draw[black] (3,1) -- (3,0);
\draw[black] (3,1) -- (4,2);
\draw[black] (4,2) -- (4,3);
\draw[black] (4,2) -- (5,2);
\draw[snake=snake,segment aspect=0]           (0.8,-1)  -- (1.2,-1);
\draw[snake=snake,segment aspect=0]           (2.8,0)  -- (3.2,0);
\draw[snake=snake,segment aspect=0]           (1.8,2)  -- (2.2,2);
\draw[snake=snake,segment aspect=0]           (3.8,3)  -- (4.2,3);
\draw[snake=snake,segment aspect=0]           (0,-0.2)  -- (0,0.2);
\draw[snake=snake,segment aspect=0]           (-0.1,-0.2)  -- (-0.1,0.2);
\draw[snake=snake,segment aspect=0]           (5,1.8)  -- (5,2.2);
\draw[snake=snake,segment aspect=0]           (5.1,1.8)  -- (5.1,2.2);
\end{tikzpicture}
\end{center}
\caption{\footnotesize Network model description of conformal blocks and Nekrasov functions \cite{network1,network2}.
Dell deformation corresponds to compactification in both horizontal and vertical directions,
the two elliptic parameters are associated with the two twists \cite{Mironov:2015thk,Mironov:2016cyq}.}
\end{figure}

In all these approaches, a full Dell theory is still a challenge
despite a considerable progress during the last years.
Recently, in \cite{AKMMdell1} we suggested to add the fourth approach to these studies:
the Shiraishi functor \cite{AKMMdell3}, which is a very explicit construction in the spirit of Noumi-Shiraishi generalization of hypergeometric series.
This approach is equivalent to studying hypergeometric functions just at the level
of series, without a reference to their free-field and representation theory
interpretations, and thus it is less involved and much simpler in many respects.
In this paper, we further elaborate along these lines,
and consider an elliptic deformation of the original Shiraishi series (a particular example of the general construction in \cite{AKMMdell3}) partly inspired by \cite{Fukuda:2020czf},
which provides an extra parameter needed for description of the Dell system.
Remarkably, this deformation (and in fact many others, see the sec.\ref{conc})
preserves the most important features of Shiraishi series,
which makes further development of this generalized hypergeometric theory
both enjoyable and potentially important for various physical applications. Note that, potentially, there could be still a problem with getting symmetric polynomials after deformation \cite{AKMMdell3}, however, it turns out that exactly the elliptic deformation of the Shiraishi series preserves symmetric polynomials.

In what follows, we study various properties of this elliptic lift of the Shiraishi functions (ELS-functions). In section 2, we define the ELS-functions and study their limit to the functions of
the dual Ruijsenaars system (when one of the elliptic parameters goes to zero; in terms of the gauge theory, this
corresponds to the perturbative limit). In section 3, we discuss the property of the ELS-functions to give rise to symmetric functions after a proper specialization. In section 4, we discuss a space-time picture behind the ELS-functions and various dualities among its parameters. In section 5, we discuss another representation of the ELS-functions very close to the
Noumi-Shiraishi one \cite{NShi}, its relation with the ordinary Shiraishi functions at a peculiar value of parameter and with the doubly compactified DIM networks. In section 6, we demonstrate that the answer for the resolved conifold for such a network calculated earlier from the refined topological vertex (DIM intertwining operator) \cite{Awata:2005fa} or, equivalently, as a trace of the product of two intertwining operators \cite{AKMMSZ} coincides with the ELS-function in the simplest $U(1)$ case. In section 7, we find an explicit
plethystic formula for this simplest ELS-function, which is a
non-trivial elliptic  generalization of the $(q,t)$ Nekrasov-Okounkov
formula from $5d$. This formula hints that the ELS-functions provide generating functions of elliptic genera of the Hilbert scheme.
In the next section 8, we demonstrate that, indeed, the ELS-function computes the generation function of the elliptic genera of the affine
Laumon spaces, which extends the claim of \cite{Shi} that
the ordinary Shiraishi function is a generating function of the Euler characteristics of the affine
Laumon spaces. These considerations lead us in section 9 to our main conjecture: that the ELS-function satisfies non-stationary equations with (yet not completely understood) quantum Dell Hamiltonians. In the Nekrasov-Shatashvili limit, it is an eigenfunction of the Dell Hamiltonians, while, in the full $\Omega$-background, it solves a non-stationary Dell equation. Section 10 contains some concluding remarks.

\section{ELS-function}

The Shiraishi function $\mathfrak{P}_n (x_i ; p \vert y_i ; s \vert q,t)$
is originally defined \cite{Shi} to be a formal power series
\beq\label{non-stationary}
\mathfrak{P}_n(x_i ; p \vert y_i ; s \vert q,t)
:= \sum_{\vec\lam} \prod_{i,j=1}^n
\frac{\mathrm{N}_{\lam^{(i)}, \lam^{(j)}}^{(j-i \vert n )} (t y_j/y_i \vert q,s)}
{\mathrm{N}_{\lam^{(i)}, \lam^{(j)}}^{(j-i \vert n )} (y_j/y_i \vert q,s)}
\prod_{\beta=1}^n \prod_{\alpha \geq 1} \left( \frac{p x_{\alpha + \beta}}{t x_{\alpha + \beta -1}} \right)^{\lam_\alpha^{(\beta)}},
\eeq
where $\vec\lam := (\lam^{(1)}, \cdots, \lam^{(n)})$ is an $n$-tuple of partitions (Young diagrams) and
$\lam_\alpha^{(\beta)}$ denotes the length of the $\alpha$-th row of the Young diagram $\lam^{(\beta)}$. We also identify $x_{\alpha+n}=x_\alpha$.
The Nekrasov factor with the $\mathrm{mod}~n$ selection rule

\beq
\mathrm{N}_{\lam, \mu}^{(k \vert n)} (u \vert q, s) :=
\prod_{j \geq i \geq 1 \atop j - i \equiv k~(\mathrm{mod}~n)} (uq^{-\mu_i + \lam_{j+1}} s^{-i +j} ;q)_{\lam_j - \lam_{ j+1}} \cdot
\prod_{\beta \geq \alpha \geq 1 \atop \beta  - \alpha \equiv -k-1~(\mathrm{mod}~n)}
 (uq^{\lam_\alpha - \m_\beta} s^{\alpha - \beta -1} ;q)_{\m_\beta - \m_{\beta+1}}, \label{Nfactor}
\eeq
with the $q$-Pochhammer symbols
\beq\label{q-shift}
(u ; q)_\infty := \prod_{i=0}^\infty ( 1- q^i u), \qquad (u ; q)_n: = \frac{( u ; q)_\infty}{(q^n u ; q)_\infty}
\eeq
is featured as the coefficients of the Shiraishi function.

We can consider an elliptic lift of $\mathfrak{P}_n (x_i ; p \vert y_i ; s \vert q,t)$
 by replacing the infinite $q$-Pochhammer symbol \eqref{q-shift} with the elliptic gamma function
 \beq
\Gamma ( u ; q,w) := \frac{(qw/u ; q , w)_\infty}{(u ; q,w)_\infty},
\eeq
where
\beq
(u ; q,w)_\infty := \prod_{i, j =0}^\infty ( 1- q^i w^j u)
= \exp \left(-\sum_{n=1}^\infty \frac{1}{n} \frac{u^n}{(1-w^n)(1-q^n)}\right).
\eeq
Note that the elliptic gamma function $\Gamma ( u; q,w)$ is symmetric in $q$ and $w$
and satisfies the following basic $q$-difference equation:
\beq
\Gamma(qu ; q,w) = \theta_w(u) \Gamma(u ; q,w),
\eeq
where our convention for the theta function\footnote{Notice an unusual normalization as compared with the standard odd $\theta$-function \cite{BatE}: the factor $(w;w)_\infty/\sqrt{z}$ is omitted.} is 
\beq
\theta_w (z) := (z ; w)_\infty (wz^{-1} ; w)_\infty.
\eeq
Since
\beq
\lim_{ w \to 0} \Gamma ( u ; q,w) = (u ; q)_\infty^{-1},
\eeq
the function
\beq\label{fracT}
\Theta(u; q,w)_n = \frac{\Gamma( q^n u ; q, w)}{\Gamma(u ; q,w)}
\eeq
has the property
\beq\label{pto0}
\lim_{ w \to 0} \Theta(u; q,w)_n  =  (u ; q)_n.
\eeq
We also have
\beq\label{Theta}
\Theta(u; q,w)_n =
\begin{cases}
\displaystyle{\prod_{k=0}^{n-1}} \theta_w (q^k u), \qquad n \geq 0, \\
\displaystyle{\prod_{k=0}^{n-1}} \theta_w (q^{-k-1} u)^{-1}, \qquad n < 0.
\end{cases}
\eeq
Now we define the elliptic lift of the Shiraishi function (ELS-function)

\bigskip

\fbox{\parbox{15cm}{
\beq
\mathfrak{P}_n^{E.G.} { (x_i ; p \vert y_i ; s \vert q,t,w)}
:= \sum_{\vec\lam} \prod_{i,j=1}^n
\frac{\mathcal{N}_{\lam^{(i)}, \lam^{(j)}}^{(j-i)} (t y_j/y_i \vert q,s,w)}
{\mathcal{N}_{\lam^{(i)}, \lam^{(j)}}^{(j-i)} (y_j/y_i \vert q,s,w)}
\prod_{\beta=1}^n \prod_{\alpha \geq 1} \left( \frac{p x_{\alpha + \beta}}{t x_{\alpha + \beta -1}} \right)^{\lam_\alpha^{(\beta)}},
\label{EG}
\eeq
where
\beqa \label{EGfactor}
\mathcal{N}_{\lam, \mu}^{(k)} (u \vert q, s, w)
&:=&
\prod_{j \geq i \geq 1 \atop j - i \equiv k~(\mathrm{mod}~n)} \Theta(uq^{-\mu_i + \lam_{j+1}} s^{-i +j} ; q,w)_{\lam_j - \lam_{ j+1}} \CR
&&~~~\times
\prod_{j \geq i \geq 1 \atop j  - i \equiv -k-1~(\mathrm{mod}~n)}
 \Theta (uq^{\lam_i - \m_j} s^{i - j -1} ; q,w)_{\m_j - \m_{j+1}}.
\eeqa}}

\bigskip

\noindent
When the selection rule is not imposed in \eqref{EGfactor}, by \eqref{fracT} we can rewrite it as
\beqa \label{ENekrasov}
\mathcal{N}_{\lam, \mu} (u \vert q, s^{-1} , w)
&=& \prod_{j \geq i \geq 1}
\frac{\Gamma(uq^{-\mu_i + \lam_{j}} s^{i - j} ; q,w)}
{\Gamma(uq^{-\mu_i + \lam_{j+1}} s^{i -j} ; q,w)}
 \prod_{i \geq j \geq 1}
\frac{ \Gamma (uq^{\lam_j - \m_{i+1}} s^{i-j +1} ; q,w)}
 {\Gamma (uq^{\lam_j - \m_i} s^{i-j+1} ; q,w)} \CR
&=& \prod_{i, j =1}^\infty \frac{\Gamma(uq^{\lam_{j} -\mu_i} s^{i-j} ; q,w)}
 {\Gamma(u s^{i-j} ; q,w)}
\cdot \frac{\Gamma (u s^{i-j+1} ; q,w)}
 {\Gamma (uq^{\lam_j - \m_{i}} s^{i-j+1} ; q,w)},
\eeqa
where we have made a change $s \to s^{-1}$ to facilitate the comparison with six dimensional gauge theory.
We see this is nothing but the elliptic lift of the Nekrasov factor
that appears in six dimensional instanton partition function \cite{Iqbal:2015fvd},\cite{Nieri:2015dts}.
The eigenfunctions of the Dell system are obtained
by taking the Nekrasov-Shatashvili limit $s \to 1$ of the Nekrasov partition function of six dimensional gauge theory
with adjoint matter and codimension two defects \cite{MMd,MM1,MM2,Koroteev:2019gqi}.
On the other hand, the original Shiraishi function
$\mathfrak{P}_n (x_i ; p \vert y_i ; s \vert q,t)$ is recovered by taking $w \to 0$.
Note that when we compare the Shiraishi function with the Nekrasov function,
not $t$ but $s$, which is introduced as a non-stationary parameter in \cite{Shi},
plays the role of the omega background parameter as it should be \cite{MMMsurfop}. This is also the case
in \cite{Braver} and \cite{BE}, where the instanton counting was discussed from the viewpoint of
affine Lie algebras. It is desirable to clarify such a double role of the parameter $s$.

To compute the limit $p \to 0$ of the function $\mathfrak{P}_n^{E.G.} { (x_i ; p \vert y_i ; s \vert q,t,w)}$,
we use
\beq\label{flip}
\Theta (u ; q, w)_n = (-u)^{n} q^{\frac{1}{2}n(n-1)} \Theta (q u^{-1} ;q, w)_{-n}^{-1}, \qquad n \in \mathbb{Z}_{\geq 0},
\eeq
which, for $n>0$,  can be proved as follows (see \eqref{Theta});
\beqa
&&\Theta (u ; q, w)_n \cdot \Theta (q u^{-1} ;q, w)_{-n}
= \prod_{k=0}^{n-1} \theta_{w} (q^k u) \theta_{w}(q^{-k}u^{-1})^{-1} \CR
&=& \prod_{k=0}^{n-1} \frac{(q^k u ;w)_\infty (wq^{-k}u^{-1} ; w)_\infty}
 {(q^{-k} u^{-1} ;w)_\infty (wq^{k}u ; w)_\infty}
 =  \prod_{k=0}^{n-1} \frac{1-q^k u}{1- q^{-k} u^{-1}} =  (-u)^{n} q^{\frac{1}{2}n(n-1)}.
\eeqa
In fact, using the relation \eqref{flip}, we can make computations similar to Appendix A of \cite{AKMMdell1}
to show that the function
\beq\label{p0limit}
Z (x_i \vert y_i  \vert q,t,w) = \lim_{p \to 0} \mathfrak{P}_n^{E.G.} { (p^{n-i}x_i ; p \vert s^{n-i}y_i ; s \vert q,t,w)}
\eeq
is the partition  function of the $S^1$ lift of the $T[U(n)]$ theory, sometimes called 4D holomorphic block \cite{Nieri:2015yia}\footnote{It would be interesting to find its relation with the twisted indices in the spirit of \cite{CDZ}.},
which is a theory on $\mathbb{R}^2 \times T^2$. The elliptic parameter $w$ is the modulus of $T^2$.
Then by the change of variables $y_i = 1/\mu_i$, we find \eqref{p0limit} agrees with
\beqa
Z (\vec{\mu}, \vec{\tau} \vert q, t, w)
=
\sum_{\{ k_i^{(a)}\}} \prod_{a=1}^{n-1} \left( t \frac{\tau_{a}} {\tau_{a+1}} \right)^{\sum_{i=1}^a k_i^{(a)}}
\prod_{i \neq j}^a \frac{\Theta\left( t \frac{\mu_i}{\mu_j} ; q, w\right)_{k_i^{(a)} - k_j^{(a)}}}
{\Theta\left( \frac{\mu_i}{\mu_j} ; q, w\right)_{k_i^{(a)} - k_j^{(a)}}}\ \prod_{i=1}^a \prod_{j=1}^{a+1} \frac{\Theta\left( \frac{q \mu_i}{t \mu_j} ; q, w\right)_{k_i^{(a)} - k_j^{(a+1)}}}
{\Theta\left( \frac{q \mu_i}{\mu_j} ; q, w\right)_{k_i^{(a)} - k_j^{(a+1)}}}\ , \label{4dblock}
\eeqa
which should be compared with eq. (4.21) of \cite{Koroteev:2019gqi}.

\section{Symmetric polynomials from ELS-functions}

As explained in \cite{AKMMdell3}, one can consider a generic construction (\ref{EG})-(\ref{EGfactor}) with the $\theta$-function in (\ref{Theta}) substituted by an almost arbitrary function $\xi(z)$ restricted only by the conditions
\be
\xi(1)=0,\ \ \ \ \ \xi(z^{-1})=\alpha z^{-1}\xi(z),\ \ \ \ \alpha\in\mathbb{C}
\ee
the corresponding generalization of the Shiraishi function being denoted through $\mathfrak{P}_n^{\xi} { (x_i ; p \vert y_i ; s \vert q,t)}$ instead of $\mathfrak{P}_n^{E.G.} { (x_i ; p \vert y_i ; s \vert q,t,w)}$ in (\ref{EG}).

Such a construction is naturally related to symmetric polynomials. That is, for a Young diagram $R$ with the line lengths $R_1\ge R_2 \ge \ldots$
\beq\label{20}
{\bf P}^\xi_R(x_i;p,s,q,t):=\prod_{i=1}^nx_i^{R_i}\cdot
\mathfrak{P}_n^{\xi} { \left(p^{n-i}x_i ; p\, \Big|\, y_i=q^{R_i}(st)^{n-i} ; s \,\Big|
\, q,\frac{q}{t}\,\right)}
\eeq
is a graded function of variables $x_i$ of the weight $|R|=\sum_iR_i$,
which is a series in $p^{nk}$:
\be
{\bf P}^\xi_R(x_i;p,s,q,t)=\sum_{k\ge 0}p^{nk}\cdot{{\bf P^\xi}^{(k)}_R(x_i;s,q,t)\over \prod_{i=1}^n x_i^k}=
\sum_{k\ge 0}{\bf P^\xi}^{(k)}_R(x_i;s,q,t)\cdot\prod_{i=1}^n\left({p\over x_i}\right)^k
\ee
Here ${\bf P^\xi}^{(k)}_R(x_i;s,q,t)$ is a polynomial of variables $x_i$ with grade $|R|+nk$. This polynomial is generally not symmetric, and we had to make a special projection in \cite{AKMMdell3} in order to produce a symmetric polynomial out of this.

This polynomial is, however, always symmetric for $\xi(z)=1-z$, and ${\bf P^\xi}^{(0)}_R(x_i;s,q,t)$ is the $n$-independent Macdonald polynomial. It turns out that the elliptic deformation $\xi(z)= \theta_w(z)$ preserves the symmetricity and $n$-independency, i.e. the ELS-function gives rise to symmetric polynomials!

For instance, in the simplest non-trivial case of $n=4$ and ${\bf P^\xi}^{(0)}_R(x_i;s,q,t)$, the symmetricity at $R=[3,1]$
requires that
\be\label{1st}
\zeta_2(1)^2-\zeta_2(1)\zeta_2(qt)-\zeta_2(1)\zeta_2(q)+\zeta_2(t)\zeta_2(qt)=0\\
\zeta_k:={\xi(q^kz)\xi(tz)\over\xi(q^{k-1}tz)\xi(qz)}
\ee
while that at $R=[4,1]$, (\ref{1st}) and also
\be
\zeta_2(1)\zeta_3(1)-\zeta_2(q^2t)\zeta_3(1)-\zeta_4(1)+\zeta_4(t)=0\label{2nd}\\
\zeta_3(1)-\zeta_2(1)\zeta_2(q^2t)-\zeta_3(q)+\zeta_3(qt)=0
\ee
etc.
These identities are, indeed, true for the theta-function, $\xi(z)= \theta_w(z)$.
The identities are four-term and each term is a product of six $\theta$-functions, they look similar to the ones appearing in description of
the elliptic ${\cal R}$-matrices in \cite{Var1,Var2}. The first identities (\ref{1st}), (\ref{2nd}) emerged even earlier in \cite[Eq.(2.31)]{BFHR} (upon identification $a^2=t^2q$ and $a^2=t^3q$ accordingly with further interchanging $q\leftrightarrow t$) within the framework of an elliptic algebra\footnote{Actually, this elliptic algebra is looking very similar to the algebra of the conjugate GNS polynomials, $\Gns^\perp$ of \cite{AKMMdell3} at $\xi(z)= \theta_w(z)$. In particular, the generalized Littlewood-Richardson coefficients do not vanish iff they do not vanish for the corresponding Schur functions, i.e. the ring of the conjugate GNS polynomials in the elliptic case is consistent with the tensor product of representations of $SL(N)$, like the algebra in \cite{BFHR} does.
We are grateful to the referee of our paper who attracted our attention to the paper \cite{BFHR}.} of surface defects in $4d$.

Note that the specialization of the ``dual" variables $y_i=q^{R_i}(st)^{n-i}$ looks similar to ``the principal specialization" \cite{Mac},\cite[Appendix B.3]{Awata:2008ed}, though the ELS-function is neither a symmetric function of these variables, nor a double-symmetric in $x_i$, $y_i$. It would be interesting to get a gauge theory interpretation for this specialization, maybe in the spirit of \cite{Kor}.

\section{Duality properties}

The ELS-function $\mathfrak{P}_n^{E.G.} { (x_i ; p \vert y_i ; s \vert q,t,w)}$
depends on five parameters\footnote{Notice that we use the letter $p$, which is an arbitrary parameter and is {\it not obligatory equal} to $q/t$ often used in the papers on the subject.}
$(p,s,q,t,w)$ other than the $PQ$ dual spectral parameters $(x_i, y_i)$.
Let us describe the geometric picture behind these parameters.

It is tempting to identify the five ELS-function parameters with $\Omega$-background of
ten dimensional space-time, for example $\mathbb{R}^4 \times {\mathrm CY}_3$,
or $\mathbb{R}^4 \times S^1 \times {\mathrm CY}_3$
for $M$ theory, where $CY_3$ is a (local) Calabi-Yau three-fold.
One can say this is cavalier, since our five parameters are independent
and we do not impose the Calabi-Yau condition.
However, it is inspired by the argument in \cite{Okounkov:2018yjl},
where the three dimensional mirror symmetry, which is an incarnation of the spectral duality
is formulated as follows:
we consider a ten dimensional space-time
\beq
Z = {\mathcal L}_4 \oplus {\mathcal L}_5 \longrightarrow X,
\eeq
where $X$ is a local curve
\beq
X = {\mathcal L}_1 \oplus {\mathcal L}_2 \longrightarrow B,
\eeq
which is a sum of two line bundles over the base curve $B$.
In the present case, $X$ is a resolved conifold and the triplet $(t,p,w)$ (see their association with gauge theory parameters in (\ref{par}) below) is
associated with $X$. Furthermore, it is natural to identify the parameters $(q,s)$,
which are the original $\Omega$-background of Nekrasov for $\mathbb{R}^4$.
Then the claim of \cite[sec.5.5]{Okounkov:2018yjl} is that geometrically
the three dimensional mirror symmetry is nothing but the exchange of
${\mathcal L}_1 \oplus {\mathcal L}_2$ and ${\mathcal L}_4 \oplus {\mathcal L}_5$.
This, in particular, implies $p \leftrightarrow s$ under the spectral duality,
and this is consistent with the conjecture in \cite{Shi}: in the limit $w\to 0$ of the ELS-function (see (\ref{non-stationary})), $(p,s)$ have been regarded as a pair of elliptic parameters exchanged by $PQ$ (or spectral) duality in \cite{Shi}.
In fact, there are a few different duality properties behind the ELS-functions.

To begin with, one of the remarkable properties in six dimensions is
that, in the instanton partition function of the theory with the adjoint hypermultiplet
of mass $m$, the omega background parameters $(q,s)$ and
the (exponentiated) mass parameter $t=e^{-m}$ appear on an equal footing\footnote{Strictly speaking we should
choose the integration contour appropriately to achieve such a symmetry.} and we can associate
plane partitions with a triple of parameters $(q,s,t)$ \cite{Mironov:2015thk}, \cite{Mironov:2016cyq}.
This is one of the trialities.

On the other hand, as we discuss in sec.\ref{u1}, when we look at the Seiberg-Witten curve of the six dimensional theory,
a pair of elliptic moduli $(p,w)$ is naturally combined with the mass parameter $t$,
to reveal $Sp(4, \mathbb{Z})$ modular symmetry \cite{Hollowood:2003cv}.
Hence, there are two trialities: the omega-background triality of $(q,s,t)$ and the elliptic parameter triality of $(p,w,t)$ (more exactly, it is a triple $(p/t,w/t,t/(qs))$).
 Since the mass parameter $t$ is common to the both trivialities above, we may connect
Nekrasov-Shatashvili limit $s \to 1$ on the omega-background side and the \lq\lq Shiraishi limit\rq\rq\ $w \to 0$ on elliptic parameter side
through their relation to the common parameter $t$. In any case,
it is an interesting challenge to work out the relation of the two limits of the ELS-function.

Note that similar trialities in five and six dimensional superconformal indices
and little string theories have been observed in \cite{LV} and \cite{BHIR}.
It is an interesting problem to figure out a relation to the triality of
$(q,s,t)$ here, since all of them are expected to be related to
the triality of DIM algebra.

\section{Shiraishi function and the compactification of DIM network}

We can similarly define an elliptic lift of the Noumi-Shiraishi
representation \cite{NShi} of the Macdonald function.
Let $M = ( m_{i,j} )_{1 \leq i,j \leq n}$ be a strictly upper triangular $n \times n$ matrix
with non-negative integer components; $m_{i,j} \in \mathbb{Z}_{\geq 0}$ and $m_{i,j}=0$ for $i \geq j$.
We define
\beqa
C_n^{\mathrm{E.G.}} (M ; y_i \vert q, t, w)
&:=& \prod_{k=2}^{n} \prod_{1 \leq i < j \leq k}
\frac{\Theta( q^{\sum_{a=k+1}^{n}(m_{i,a} - m_{j,a})} \frac{t y_j}{y_i} ; q, w)_{m_{i,k}}}
{\Theta( q^{\sum_{a=k+1}^{n}(m_{i,a} - m_{j,a})}\frac{q y_j}{y_i}; q, w)_{m_{i,k}}}\times
\CR
&\times&\prod_{1 \leq i \leq j < k} \frac{\Theta(q^{- m_{j,k} + \sum_{a=k+1}^{N}(m_{i,a} - m_{j,a})}
 \frac{q y_j}{t y_i} ; q, p)_{m_{i,k}}}
{\Theta(q^{- m_{j,k} +\sum_{a=k+1}^{n}(m_{i,a} - m_{j,a})} \frac{y_j}{y_i} ; q, w)_{m_{i,k}}}.
\eeqa
Then the elliptic gamma lift $f^{\mathrm{E.G.}}_{n} (x \vert y \vert q, t, p)$ is defined by
\beq
f^{\mathrm{E.G.}}_{n} (x_i \vert y_i \vert q, t, w) := \sum_{m_{i,j}} C_n^{\mathrm{E.G.}} (M ; y_i \vert q, \frac{q}{t}, w)
\prod_{1 \leq i<j \leq n} \left( \frac{x_j}{x_i} \right)^{m_{i,j}}. \label{EllipticGamma}
\eeq
Note that we have made a change of variable from $t$ to $q/t$. In fact,
$f^{\mathrm{E.G.}}_{n} (x \vert y \vert q, t, w)$ is equal \cite{AKMMdell3} to
\beq
f^{\mathrm{E.G.}}_{n} (x \vert y \vert q, t, w)=\lim_{p \to 0} \mathfrak{P}_n^{E.G.} { (p^{n-i}x_i ; p \vert s^{n-i}y_i ; s \vert q,t,w)}=Z (x_i \vert y_i  \vert q,t,w)
\eeq
By the relation \eqref{pto0}, $f^{\mathrm{E.G.}}_{n} (x_i \vert y_i \vert q, t, w)$ reduces to the Noumi-Shiraishi
representation of the Macdonald function in the limit $w \to 0$.
%
%
%
%

One of the main results of a recent paper \cite{Fukuda:2020czf}
is that up to some normalization constant $\mathfrak{C}$,
the elliptic lift $f^{\mathrm{E.G.}}_{n} (x_i \vert y_i \vert q, t, p)$ gives
the $PQ$ dual form of the Shiraishi function $\mathfrak{P}_n(x_i;p|y_i;s|q,t)$
{\it on the special locus $s = t^{-1/n}$}. Namely they proved that
\beq\label{FOSid}
\mathfrak{P}_n (\tilde{x}_i; p^{1/n}  \vert \tilde{y}_i; t^{-1/n} \vert q,t)
= \mathfrak{C} \cdot f^{\mathrm{E.G.}}_{n} (y_i \vert x_i \vert q, t, p),
\eeq
where the spectral parameters  $x_i$ and $y_i$ are scaled as
\beq
\tilde{x}_k := p^{-k/n} x_k, \qquad \tilde{y}_k := t^{k/n} y_k.
\eeq
The normalization constant is
\beq
\mathfrak{C} :=
\left( \frac{(qp/t ; q, p)_\infty}{(p;p)_\infty (pt; q, p)_\infty} \right)^n
\prod_{1 \leq i < j \leq n} \frac{\Gamma(t x_i/x_j; q, p)} {\Gamma(q x_i/x_j; q, p)}
\prod_{1 \leq i < j \leq n} \frac{(t x_i/x_j; q)_\infty} {(q x_i/x_j; q)_\infty}.
\eeq

Originally $p$ and $s$ are a pair of elliptic parameters in the Shiraishi function, but on the r.h.s. of
\eqref{FOSid} one of the parameters is constrained by $s= t^{-1/n}$, and
the remaining parameter $p$ appears as the elliptic modulus of the theta function.
Note that the variables $x_i$ and $y_i$ are exchanged on the left and the right hand sides of \eqref{FOSid}.
In this sense, $f^{\mathrm{E.G.}}_{n}$ gives a $PQ$ dual representation of the Shiraishi function.
The restriction to the special locus $s = t^{-1/n}$ comes from the fact that
the affine screening operators employed in \cite{Shi} can be reproduced from the intertwiners
of the Ding-Iohara-Miki (DIM) algebra only when $s = t^{-1/n}$ \cite{Fukuda:2020czf}.
In the proof of \eqref{FOSid}, the duality formula for the DIM intertwiners established in \cite{Fukuda:2019ywe}
was used, and this is the reason why it is valid only on the special locus.
That is, \eqref{FOSid} is a consequence of equating the results of computing the same network diagram
in two different ways exchanging the vertical and the horizontal direction.
The fact that the affine screening operators in \cite{Shi} agree with the DIM screening operators
only at $s = t^{-1/n}$ seems consistent with the fact that the DIM algebra does not have any elliptic
parameter. We can extend DIM algebra to an elliptic DIM algebra with the elliptic parameter $w$
(see the next section).
However, the affine screening operators in \cite{Shi} suggests the existence of another elliptic
deformation with parameter $s$.

The computation of \cite{Fukuda:2020czf} should be compared with that of the M-string partition function
\cite{Haghighat:2013gba}, \cite{Haghighat:2013tka}.
The difference is that there are no surface defects in \cite{Haghighat:2013gba}, \cite{Haghighat:2013tka},
and we would obtain the formulas without the mod $n$ selection rule.
The M-string partition function is obtained from the toric (network) diagram
where external legs are identified (compactified) along
either of the vertical (preferred) and the horizontal (unpreferred) directions.
Hence, there are two ways of computing the M-string partition function, which are related by duality.
Comparing the formulas (Case I and Case II in section 3 of \cite{Haghighat:2013gba}),
we find that the compactification along the vertical direction corresponds to the original definition of
the Shiraishi function, while the compactification along the horizontal direction gives
the elliptic lift $f^{\mathrm{E.G.}}_{n}$.

\section{ELS-function and the double compactified DIM network\label{u1}}

Once we understand that the Shiraishi function comes from the network with one direction compactified,
it is natural  to expect that the network with double compactification gives the ELS-function
$\mathfrak{P}_n^{E.G.} { (x_i ; p \vert y_i ; s \vert q,t,w)}$.
The relation of double compactified toric diagram to the Dell system has been expected since \cite{Hollowood:2003cv}.
In the simplest case ($n=1$), the corresponding diagram represents the resolved conifold geometry.
The single compactification produces a (degenerate) genus one curve
as a pair of rational curves $\mathbb{P}^1$ attached together
at two punctures,  while the double compactification gives a (degenerate) genus two curve consisting of
a triple of rational curves attached together at two punctures  (see Fig. 2).
Note that the genus two curve has three moduli, which are two elliptic moduli $p$ and $w$
and the mass of the adjoint matter $t=e^{-m}$.
Since these parameters come from the K\"ahler parameters of three rational curves $\mathbb{P}^1$,
we may expect a triality among them.

\begin{figure}[h]
 \begin{center}
 \includegraphics[width=10cm,clip]{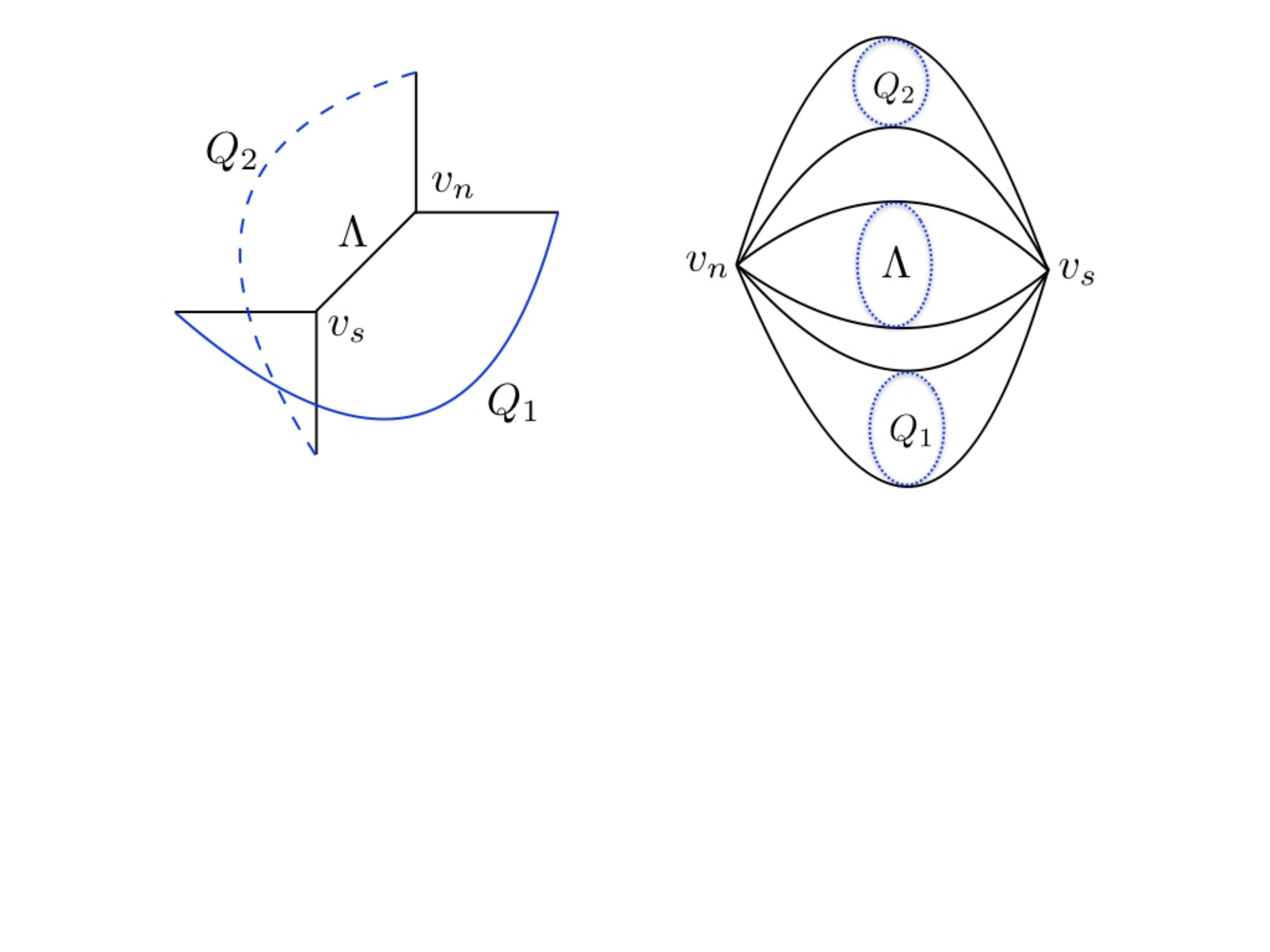}
 \end{center}
\vspace{-3.5cm}
 \caption{\footnotesize Double compactification of the resolved conifold is a degenerate genus two curve.
 The K\"ahler parameters of three rational curves $\mathbb{P}^1$ match three moduli
 of genus two curves. }
\end{figure}

In \cite{Awata:2005fa} the (refined) elliptic genus of $\mathrm{Hilb}_n \mathbb{C}^2$ was computed by the method of
the refined topological vertex. The relevant toric diagram is the double compactified conifold diagram.
To the diagram, we can associate the K\"ahler parameter $\Lambda$ of the base $\mathbb{P}^1$
and those of the compactified two dimensional fiber $Q_1$ and $Q_2$.
Since the conifold geometry engineers $U(1)$ gauge theory, the fixed points of the torus action are
labeled by a single Young diagram $\lambda$.
The result of \cite{Awata:2005fa} is
\beqa\label{AK}
Z_{U(1)}^{6d}(Q_i, \Lambda ; q,t)  &=& \sum_{\lambda} (\gq \Lambda)^{|\lambda|} \prod_{k=0}^{\infty} \prod_{\square \in \lambda}
\frac{(1- \gq^{-1} Q_1^{k+1} Q_2^k q^{a(\square)} t^{\ell(\square) +1})(1- \gq^{-1} Q_1^{k} Q_2^{k+1} q^{a(\square)} t^{\ell(\square)+1})}
{(1- Q_1^{k} Q_2^k q^{a(\square)} t^{\ell(\square) +1})(1- \gq^{-2} Q_1^{k+1}Q_2^{k+1} q^{a(\square)} t^{\ell(\square)+1})} \CR
&&~~~\times \frac{(1- \gq^{-1} Q_1^{k+1} Q_2^k q^{-a(\square)-1} t^{-\ell(\square)})(1- \gq^{-1} Q_1^{k} Q_2^{k+1} q^{-a(\square)-1} t^{-\ell(\square)})}
{(1- Q_1^{k} Q_2^k q^{-a(\square)-1} t^{-\ell(\square)})(1- \gq^{-2} Q_1^{k+1}Q_2^{k+1} q^{-a(\square)-1}
t^{-\ell(\square)})}
\eeqa
where $\gq = (t/q)^{1/2}$ and, for $\square=(i,j)$, we define the arm length $a(\square) := \lam_i - j$
and the leg length $\ell(\square) := \lam^\vee_j -i$, with $^\vee$ denoting the transposition of the Young diagram.
Making the change of variables
\beq\label{transformation}
w:= Q_1 Q_2, \qquad s:=t^{-1},
\eeq
we can rewrite the above formula in terms of the theta functions:
\beq
Z_{U(1)}^{6d} (w,Q_1, \Lambda ; q,s)  =
 \sum_{\lambda} (\gq \Lambda)^{|\lambda|}
 \prod_{\square \in \lambda}
\frac{\theta_w (\gq^{-1} Q_1 q^{a(\square)} s^{-\ell(\square)-1}) \theta_w (\gq^{-1} Q_1 q^{-a(\square)-1} s^{\ell(\square)})}
{\theta_w (q^{a(\square)} s^{-\ell(\square)-1}) \theta_w (q^{-a(\square)-1} s^{\ell(\square)})}.
\eeq

Note that using the formula proved in \cite{Awata:2008ed},
we can show the following identity
for the elliptic Nekrasov factor \cite{Nieri:2015dts};
\beqa\label{eNek}
\mathcal{N}_{\lam, \mu} (u \vert q, s^{-1} , w)
&=& \prod_{i, j =1}^\infty \frac{\Gamma(uq^{\lam_{j} -\mu_i} s^{i-j} ; q,w)}
 {\Gamma(u s^{i-j} ; q,w)}
\cdot \frac{\Gamma (u s^{i-j+1} ; q,w)}
 {\Gamma (uq^{\lam_j - \m_{i}} s^{i-j+1} ; q,w)} \CR
 &=&
\prod_{\square \in \lam} \theta_w (u q^{a_\lambda(\square)}  s^{\ell_\mu(\square)+1})
\prod_{\square \in \mu} \theta_w (u q^{-a_\mu(\square) -1}  s^{-\ell_\lambda(\square)}).
\eeqa
Let us look at the case $n=1$, though it is trivial as an integrable system.
In this case, the selection rule is empty, and the ELS-function becomes
a summation over a single Young diagram $\lambda$.
Using the formula \eqref{eNek}, we obtain
\beq\label{ESLn1}
\mathfrak{P}_{n=1}^{E.G.} { (p \vert s \vert q,t,w)}
= \sum_{\lam} \left( \frac{p}{t}\right)^{|\lam|}
\prod_{\square \in \lam} \frac
{\theta_w (t q^{a(\square)}  s^{-\ell(\square)-1})
\cdot \theta_w (t q^{-a(\square) -1}  s^{\ell(\square)})}
{\theta_w (q^{a(\square)}  s^{-\ell(\square)-1})
\cdot \theta_w (q^{-a(\square) -1}  s^{\ell(\square)})}.
\eeq
Hence, by the identification $\gq^{-1} Q_1=t$,  $\gq \Lambda = p/t$, we see the agreement of
$Z_{U(1)}^{6d} (w,y, \Lambda ; q,t)$ and $\mathfrak{P}_{n=1}^{E.G.} { (p \vert s \vert q,t,w)}$.

Note that equivalently one can describe the double compactification of the network as sums of traces of
two intertwiners: upon identification $p\to P_\perp$, $w\to Q$, $t\to P\sqrt{q/t}$, $s\to t^{-1}$ (notice also a rescaling of the mass parameter $m\to -2\pi im$), formula (\ref{ESLn1}) reduces to \cite[(6.31)]{AKMMSZ} obtained this way. One may try to consider an immediate generalization \cite[(6.43)]{AKMMSZ} with $2n$ intertwiners. The answer can not literally give the ELS-function, because it describes the case of $n$ variables $y_i$ and still only one
variable $x$, while in (\ref{EG}) there are both $n$ variables $y_i$ and $n$ variables $x_i$. However, if one impose all $x_i$ coinciding, the answer is ELS-function without the selection rule. In order to get the selection rule, one has to make the next step. We return to the network picture of the ELS-function elsewhere \cite{ELSDIM}.

In \cite{Hollowood:2003cv} the following symmetry among the moduli parameters
in the Seiberg-Witten curve was pointed out\footnote{We denote
the dimensionless mass parameter $\beta m$ in \cite{Hollowood:2003cv} by $-m$. The sign of the mass parameter changes under the flop of local Calabi-Yau geometry.}
\beq\label{dictionary}
\hat\tau + \frac{m}{2\pi i}  \longleftrightarrow \tau + \frac{m}{2\pi i}
 \longleftrightarrow \frac{m}{2\pi i}
\eeq
In terms of the exponentiated parameters
\beq\label{par}
w = e^{2 \pi i \hat\tau}, \qquad p = e^{2 \pi i \tau}, \qquad t = e^{-m},
\eeq
this is a symmetry among $(w/t, p/t, t/(qs))$ ($qs=1$ in the unrefined case).
If we take into account the relations \eqref{transformation}
to the parameters of the K\"ahler parameters of three
rational curves in the double compactified network, we can see that this symmetry comes from
the symmetry among $(\Lambda, Q_1, Q_2)$.

Let us note that (\ref{AK}) is evidently symmetric w.r.t. the permutation $Q_1\leftrightarrow Q_2$. In terms of ELS-function parameters, it is the symmetry under $w/t\leftrightarrow t/(qs)$ at constant $s$ and $t$. Let us test the second, less evident $p/t\leftrightarrow t/(qs)$ symmetry, which one would expect because of the duality between the elliptic parameters $p$ and $w$. Note that the symmetry $p/t\leftrightarrow t/(qs)$ can be checked already at the level of the ordinary Shiraishi function. To do this, we consider the $w=0$ limit of (\ref{AK}).

In the limit of $Q_2=0$, formula (\ref{AK}) reduces to
\beq\label{tneq}
Z_{U(1)}^{5d} (Q_1, \Lambda ; q,t)  =
 \sum_{\lambda} (\gq \Lambda)^{|\lambda|}
 \prod_{\square \in \lambda}
\frac{(1-\gq^{-1} Q_1 q^{a(\square)} t^{\ell(\square)+1})(\gq^{-1} Q_1 -q^{a(\square)+1} t^{\ell(\square)})}
{ (1-q^{a(\square)} t^{\ell(\square)+1}) (1-q^{a(\square)+1} t^{\ell(\square)}) }.
\eeq
This is the Shiraishi function at $n=1$ (upon an evident identification of parameters) \cite{Shi}. The symmetry we check is $\Lambda\leftrightarrow Q_1$ at constant $q$ and $t$.

Let us first put $t=q$ and note that $a(\square)+\ell(\square)+1=h_\square$, the hook length. Then,
\beq
Z_{U(1)}^{5d} (Q_1, \Lambda ; q)  =
 \sum_{\lambda} \Lambda^{|\lambda|}
 \prod_{\square \in \lambda}
\frac{(1-Q_1 q^{h_\square})(Q_1 -q^{h_\square})}
{ (1-q^{h_\square}) ^2 }.
\eeq
One can check that this formula is not symmetric w.r.t. the permutation of $\Lambda$ and $Q_1$: it is possible to perform the summation explicitly \cite[Eq.(33)]{Ch}\footnote{Notice various misprints in this equation in that paper.},
\be
Z_{U(1)}^{5d} (Q_1, \Lambda ; q)
=\!\! \prod_{k,m\ge 0}\left[{(1-\Lambda^{k+1}Q_1^{k}q^{m+1})(1-\Lambda^{k}Q_1^{k+1}q^{m+1})\over
(1-\Lambda^{k+1}Q_1^{k+1}q^{m+1})(1-\Lambda^{k+1}Q_1^{k+1}q^{m+1}) }\right]^{m+1}
\!\!\!\!\!\!\!\!\cdot \prod_{k=1}{1\over (1-\Lambda^k Q_1^k)}\cdot
\prod_{m=1}{1\over (1-Q_1q^m)^m}
\nn
\ee
and what spoils the symmetry is the last factor. Fully $(\Lambda,Q_1 )$-symmetric is the product
\beq\label{M}
\prod_{m=1}(1-Q_1q^m)^m\cdot Z_{U(1)}^{5d} (Q_1, \Lambda ; q)
\eeq
The formula for $Z_{U(1)}^{5d} (Q_1, \Lambda ; q)$ can be rewritten in terms of the plethystic exponential
\be
\hbox{P.E.}\Big[F(\{x_i\})\Big]:=\exp\left(\sum_k{1\over k}F(\{x_i^k\})\right)
\ee
where by $\{x_i\}$ we understand all the variables $(Q_i, \Lambda, q,t)$, and
\be
Z_{U(1)}^{5d} (Q_1, \Lambda ; q)=\hbox{P.E.}\Big[{\Lambda (1-qQ_1)(Q_1-q)\over (1-\Lambda Q_1)(1-q)^2}\Big]=
\hbox{P.E.}\Big[{\Lambda Q_1(1+q^2)-q(\Lambda+Q_1)\over (1-\Lambda Q_1)(1-q)^2}+{qQ_1\over (1-q)^2}\Big]
\ee
The last term is the only non-symmetric one, and it exactly corresponds to the MacMahon function as in (\ref{M}).

At last, at the generic values of $t$, (\ref{tneq}), one similarly obtains \cite[Eq.(1.4)]{AK09} (this formula was first discussed in \cite{Iqbal:2008ra,Poghossian:2008ge,Awata:2008ed})
\beq
Z_{U(1)}^{5d} (Q_1, \Lambda ; q,t) =\hbox{P.E.}\Big[{\gq\Lambda (1-t\gq^{-1}Q_1)(\gq^{-1}Q_1-q)\over (1-\Lambda Q_1)(1-q)(1-t)}\Big]=\hbox{P.E.}\Big[{\Lambda Q_1(1+tq)-\sqrt{qt}(\Lambda+Q_1)\over (1-\Lambda Q_1)(1-q)(1-t)}+
{\sqrt{qt}Q_1\over (1-q)(1-t)}\Big]
\eeq
Again, only the last term is non-symmetric, and this means the invariant is the product
\beq
\prod_{l,m\ge 0}(1-Q_1q^{l+{1\over 2}}t^{m+{1\over 2}})\cdot Z_{U(1)}^{5d} (Q_1, \Lambda ; q,t)
\eeq
This should not come as a surprise: $Z_{U(1)}^{5d} (Q_1, \Lambda ; q,t)$ describes the instanton sum in the $5d$ theory with adjoint matter hypermultiplet, while this additional factor is the perturbative contribution \cite{Awata:2005fa}, and the full partition function is invariant. This factor also coincides with the normalization factor in \cite[Eq.(6.31)]{AKMMSZ}.

Similarly, the perturbative factor in the $6d$ theory is given by \cite[Eq.(3.40)]{Awata:2005fa}
\beq\label{pert6}
\hbox{P.E.}\Big[-{\sqrt{qt}(Q_1+Q_2)-(1+qt)Q_1Q_2\over (1-q)(1-t)(1-Q_1Q_2)}\Big]
\eeq
Indeed, one can check (see the next section) that, in the case of general parameters, the triality symmetric function is given by
$$
\hbox{P.E.}\Big[-{\sqrt{qt}(Q_1+Q_2)-(1+qt)Q_1Q_2\over (1-q)(1-t)(1-Q_1Q_2)}\Big]\times\  Z_{U(1)}^{6d}(Q_i, \Lambda ; q,t)=
$$
$$
\prod_{i_1,i_2,i_3\ge 0}(1-Q_1^{i_1+1}Q_2^{i_1}q^{i_2+{1\over 2}}t^{i_3+{1\over 2}})
(1-Q_1^{i_1}Q_2^{i_1+1}q^{i_2+{1\over 2}}t^{i_3+{1\over 2}})
(1-Q_1^{i_1+1}Q_2^{i_2+1}q^{i_2}t^{i_3}) (1-Q_1^{i_1+1}Q_2^{i_1+1}q^{i_2+1}t^{i_3+1})\times
$$
\beq
\times\  Z_{U(1)}^{6d}(Q_i, \Lambda ; q,t)
\eeq
Similar computations for $n >1$ is more demanding, but it is interesting to see how
the above symmetry among $(p,w,t)$ survives for $n >1$.

\section{Summing up $Z_{U(1)}^{6d}(Q_i, \Lambda ; q,t)$}

One can perform a summation in formula (\ref{AK}) and present $Z_{U(1)}^{6d}(Q_i, \Lambda ; q,t)$ as a plethystic exponential. The coefficients of this exponential are expressed through a single periodic function. Indeed, let us rewrite (\ref{AK}) in the form
\beq\label{AK1}
Z_{U(1)}^{6d}(Q_i, \Lambda ; q,t)
= \sum_{\lambda} (\gq\Lambda)^{|\lambda|} \prod_{\square \in \lambda}\Phi(\gq^{-1} Q_1,Q_1Q_2;q^{a(\square)} t^{\ell(\square)}),
\eeq
where
\beq\label{Phi}
\Phi (Q,w;u):=\frac{\theta_w (utQ) \theta_w (Q/(qu))}
{\theta_w (ut) \theta_w (1/(qu))}
\eeq
The function $\Phi(\gq^{-1} Q_1,Q_1Q_2;u)$ is symmetric in $Q_1$, $Q_2$. Hence, $Z_{U(1)}^{6d}(Q_i, \Lambda ; q,t)$ celebrates the same property. We construct it in a simple way out of  $\Phi (Q,w;1)$ below.

As before, let us first consider the case of $t=q$ so that
\beq\label{AKq}
Z_{U(1)}^{6d}(Q_i, \Lambda ; q)
= \sum_{\lambda} \Lambda^{|\lambda|} \prod_{\square \in \lambda}\Phi(Q_1,Q_1Q_2;q^{h(\square)}),
\eeq
and expand the function $\Phi (Q_1,Q_1Q_2;1)$
\ (i.e. the $\Lambda$-linear term in $Z_{U(1)}^{6d}(Q_i, \Lambda ; q)$, which corresponds to the item with a single-box Young diagram $\Lambda=[1]$) \
into the power series
\beq\label{Phic}
\Phi (Q_1,Q_1Q_2;1)={q\over (1-q)^2}\cdot \sum_{i,j\ge 0}c_{i,j}^{(q)}Q_1^iQ_2^j,
\eeq
i.e.
\beq
\prod_{k=0}^\infty \frac{\Big(1-qQ_1^{k+1}Q_2^{k}\Big) \Big(1-q^{-1}Q_1^{k+1}Q_2^{k}\Big)
\Big(1-qQ_1^{k}Q_2^{k+1}\Big)\Big(1-\frac{1}{q}Q_1^{k}Q_2^{k+1}\Big)}
{\Big(1-qQ_1^{k+1}Q_2^{k+1}\Big)^2\Big(1-q^{-1}Q_1^{k+1}Q_2^{k+1}\Big)^2}
=\sum_{i,j\ge 0}c_{i,j}^{(q)}Q_1^iQ_2^j
\eeq
The coefficients $c_{i,j}^{(q)}=c_{j,i}^{(q)}$, since this function is symmetric. These coefficients sometimes vanish: they vanish within the area with the upper boundary
\be
\begin{array}{c|cccccccccccccc}
i & 0 & 1 & 2 & 3 & 4 & 5 & 6 & 7 & 8 & 9 & 10 &  \ldots \\
j & 3 & 5 & 7 & 8 & 10 & 11 & 13 & 14 & 15 & 17 & 18 & \ldots
\end{array}
\ee
while the lower boundary is given by the switch $i\leftrightarrow j$.

Since the $\theta$-function is quasi-periodic: $\theta_w(wz)=-z^{-1}\theta_w(z)$, such is the function $\Phi (Q_1,Q_1Q_2;1)$, (\ref{Phic}), and the coefficients $c_{i,j}^{(q)}$ satisfy the relation
\be
c_{i,j}^{(q)}=c_{2i-j+2,i}^{(q)}
\ee
and, being repeated $k$ times,
\be\label{per}
c_{i,j}^{(q)}=c_{i(k+1)-jk+k(k+1),ik-j(k-1)+k(k-1)}^{(q)}
\ee
Since $c_{i,j}^{(q)}\ne 0$ only at $i,j\ge 1$, this generates the vanishing coefficients $c_{i,j}^{(q)}$,

\bigskip

Now we are ready to formulate the answer for $Z_{U(1)}^{6d}(Q_i, \Lambda ; q)$: it is given by the plethystic exponential
\beq\label{AK2}
Z_{U(1)}^{6d}(Q_i, \Lambda ; q)=\hbox{P.E.}\left[{q\over (1-q)^2}\left\{F(\Lambda,Q_1,Q_2;q)
+\Lambda-{\Lambda (Q_1-q)(Q_1-q^{-1})\over 1-\Lambda Q_1}-{\Lambda (Q_2-q)(Q_2-q^{-1})\over 1-\Lambda Q_2}\right\}\right]
\eeq
where the function $F(\Lambda,Q_1,Q_2;q)$ is a symmetric function of all three variables $\Lambda$, $Q_1$ and $Q_2$ given by a power series
\beq\label{Ff}
F(\Lambda,Q_1,Q_2;q)=\sum_{i,j,k\ge 1}c_{ik,j+(i-1)(k-1)}^{(q)}\Lambda^iQ_1^jQ_2^k
\eeq
Note that, while symmetricity of the $F$-function (\ref{Ff}) w.r.t. permuting $\Lambda$ and $Q_2$ is evident from (\ref{Ff}), that w.r.t. permuting $Q_1$ and $Q_2$ is not evident, and is an additional property of the coefficients $c_{i,j}^{(q)}$.
The triple symmetry can be made explicit, if one notes that actually
\be
c^{(q)}_{ik, j+(i-1)(k-1)} = c^{(q)}_{\frac{ijk}{M} ,\,\frac{ijk}{M}+2M+1-i-j-k}
\ee
where $M_{i,j,k}:= {\rm max}(i,j,k)$, and $\frac{ijk}{M} = {\rm min}(ij, ik, jk)$,
i.e. is fully symmetric. This formula looks like tropical, which adds to mysteries of the ELS-functions, which will be discussed elsewhere..
Note also that $Z_{U(1)}^{6d}(Q_i, \Lambda ; q)$ becomes symmetric in all three variables upon multiplying (\ref{AK2}) with the perturbative factor (\ref{pert6}) at $t=q$.

\bigskip

Finally, let us consider the case of generic $t$ and again expand the function $\Phi (\gq Q_1,Q_1Q_2;t)$, i.e. the $\Lambda$-linear term in $Z_{U(1)}^{6d}(Q_i, \Lambda ; q,t)$ corresponding to the one-box Young diagram $[1]$, into the power series
\beq
\Phi (\gq^{-1} Q_1,Q_1Q_2;1)=\left.\frac{\theta_w (Q_1t) \theta_w (Q_1q^{-1})}
{\theta_w (t) \theta_w (q^{-1})}\right|_{w=Q_1Q_2}={\sqrt{qt}\over (1-q)(1-t)}\cdot\sum_{i,j\ge 0}c_{i,j}^{(q,t)}Q_1^iQ_2^j
\eeq
Again the coefficients $c_{i,j}^{(q,t)}=c_{j,i}^{(q,t)}$, since this function is still symmetric. In fact, $c_{i,j}^{(q,t)}$ have the same symmetry and vanishing properties as $c_{i,j}^{(q)}$ which we considered above. In particular, the periodicity condition leads to the relation
\be
c_{i,j}^{(q,t)}=\gq^{2k(j-i)-2k^2}\cdot c_{i(k+1)-jk+k(k+1),ik-j(k-1)+k(k-1)}^{(q,t)},\ \ \ \ \ \forall k
\ee
which differs from (\ref{per}) only by a factor of $\gq$.

Now we are ready to formulate the final answer for $Z_{U(1)}^{6d}(Q_i, \Lambda ; q,t)$: it is given by the plethystic exponential

\bigskip

\hspace{-.6cm}\fbox{\parbox{17cm}{
\beqa\label{AK3}
&&Z_{U(1)}^{6d}(Q_i, \Lambda ; q,t)=
\CR
&=&\hbox{P.E.}\left[{\sqrt{qt}\over (1-q)(1-t)}\left\{F(\Lambda,Q_1,Q_2;q,t)
+\Lambda-{\Lambda (Q_1-\sqrt{qt})(Q_1-{1\over\sqrt{qt}})\over 1-\Lambda Q_1}-{\Lambda (Q_2-\sqrt{qt})(Q_2-{1\over\sqrt{qt}})
\over 1-\Lambda Q_2}\right\}\right]\CR
\eeqa
\hspace{1cm}\parbox{15cm}{where the function $F(\Lambda,Q_1,Q_2;q,t)$ is a symmetric function of all three variables $\Lambda$, $Q_1$ and $Q_2$ given by a power series}
\beq\label{Ff1}
F(\Lambda,Q_1,Q_2;q,t)=\gq\cdot\sum_{i,j,k\ge 1}c_{ik,j+(i-1)(k-1)}^{(q,t)}\Lambda^iQ_1^jQ_2^k
\eeq
}}

\bigskip

Again, $Z_{U(1)}^{6d}(Q_i, \Lambda ; q,t)$ becomes symmetric in all three variables upon multiplying (\ref{AK3}) with the perturbative factor (\ref{pert6}).

In fact, it is highly non-trivial fact that the sum over the Young diagrams is given by a reasonable product formula (by the plethystic exponential of a simple sum).
In the $5d$ case, it was first noticed in \cite{Ch,Iqbal:2008ra,Poghossian:2008ge,Awata:2008ed} (see also later discussions in \cite{Iqbal} and \cite[sec.3.4]{Fukuda:2020czf}). Later, this kind of formula was discussed in mathematical literature \cite{CNO,RW}, however, the issue has not become clearer.  We demonstrated that the sum over the Young diagrams can be still rewritten in a relatively simple plethystic form also in the $6d$ case, with a triality emerging (as a symmetry between $\Lambda$, $Q_1$ and $Q_2$). Since this case is elliptic,  for mathematicians the formula can be probably also named  ``elliptic deformation of the $(q,t)$ Nekrasov-Okounkov formula".

Mathematically, we obtained that $Z_{U(1)}^{6d}(Q_i, \Lambda ; q,t)$ turns out to be a generating function of the elliptic genera of the Hilbert scheme of points on $\mathbb{C}^2$. Indeed, in \cite{DMVV}, a formula for the elliptic genera of the Hilbert scheme, which is a resolution of the orbifold singularities of the symmetric products, was derived from
the data of the elliptic genus of $\mathbb{C}^2$. This formula is a non-equivariant version, i.e. that without the
$\Omega$-background. Later, the formula was generalized to the equivariant version in \cite{Waelder}.

To be more concrete, since the equivariant $\chi_y$
genus is \cite[Eq. (2.25)]{AK09}
\beq
\chi_Q(\mathbb{C}^2; t_1,  t_2) = \frac{(1-Qt_1)(1-Qt_2)}{(1-t_1)(1-t_2)},
\eeq
the  equivariant elliptic genus of $\mathbb{C}^2$ is
\beq
Ell(\mathbb{C}^2;w,Q, t_1, t_2) = \frac{\theta_w(Qt_1) \theta_w(Qt_2)}{\theta_w(t_1)\theta_w(t_2)}=\Phi(Q,w;1)
\eeq
where $t_1=t$, $t_2=q^{-1}$ are equivariant parameters.

Now the equivariant version of the formula due to \cite{DMVV} says \cite[Theorem 11]{Waelder} that, if
\beq
Ell(\mathbb{C}^2;w,Q, t_1, t_2) = \sum_{m, \ell, k_1, k_2} C(m, \ell, k_i) w^m Q^\ell t_1^{k_1} t_2^{k_2}
\eeq
then
\beq
\sum_n Ell((\mathbb{C}^2)^{[n]} ;w,Q, t_1, t_2)\cdot p^n = \prod_{m,n,\ell, k_i}
\frac{1}{(1- p^n w^m Q^\ell t_1^{k_1} t_2^{k_2})^{C(nm, \ell, k_i)}}.
\eeq
After a proper choice of variables in the generating function (and a slight redefinition of this later), this formula reduces to the plethystic exponential of (\ref{Ff1}), and the r.h.s. of (\ref{AK2}), (\ref{AK3}) is, indeed, a generating function of the elliptic genera of the Hilbert scheme of points on $\mathbb{C}^2$.

Thus, formula (\ref{Ff1}) for the $F$-function means that the ELS-function, at least, at $n=1$, provides a generating function  of the elliptic genera of the Hilbert scheme of points on $\mathbb{C}^2$. We discuss this issue in the next section in detail in the case of ELS-function at general $n$.

\section{Elliptic genus of the affine Laumon space}

In this section, we argue that the ELS-function
computes the generation function of the elliptic genera of the affine
Laumon spaces.  Laumon moduli spaces are certain smooth closures of the moduli spaces of maps
from the projective line to the flag variety of $GL_n$.
In \cite{Shi}, it was shown that
the Shiraishi function agrees with the generating function of the Euler characteristics of the affine
Laumon spaces. The relation of the Calogero-Moser system
and the (non-affine) Laumon space was established by A. Negut \cite{Neg}. (See also \cite{BFS}.) An extension to the affine Laumon space see in \cite{Neg2}.

Let $X$ be a K\"ahler manifold of complex dimension $d$,
and $E$ an elliptic curve with modulus $\tau$.
Recall that a line bundle over $E$ is labeled by a parameter $z \in \mathrm{Jac}~E \simeq E$.
The elliptic genus $Ell(X)$ is defined as the genus one
(or one-loop in the sense of string theory) partition function
of $\mathcal{N}=2$ supersymmetric sigma model on $E$ with the target space $X$
\cite{Schellekens:1986yi},\cite{Witten:1986bf},\cite{Witten86}.
We impose the periodic boundary conditions on both the left- and the right-moving fermions.
This leads to the insertion of $(-1)^F$ in the trace, where
$F = F_L + F_R$ is the sum of the left- and the right-moving fermion numbers.
Using the trace over the Hilbert space of the Ramond-Ramond sector, the elliptic genus is
\beq\label{sigmamodel}
Ell(X; q,y)  := \mathrm{Tr}_{\mathcal{H}(X)}
\left[ (-1)^{F} Q^{F_L} w^{L_0 - \frac{d}{8}} \bar{w}^{\bar{L_0} - \frac{d}{8}} \right],
\eeq
where $w=e^{2\pi i \tau},~Q=e^{2\pi i z}$ and $L_0, \bar{L_0}$ are zero modes of the Virasoro algebra.
It is known that only the ground states contribute to the trace in the right moving sector \cite{Dijkgraaf:1996xw}.
When there is a torus action on the target space $X$, the elliptic genus becomes character valued.
For example, one can insert $\exp \left(\sum_{i=1}^n \epsilon_i T^i \right)$
in the trace to define an equivariant version of the elliptic genus,
where $T^{i}$ are generators of the torus action. Then the elliptic genus is a polynomial
in the exponentiated equivariant parameters $e^{\epsilon_i}$.

The elliptic genus is also defined as the index of the $\bar\partial$-operator\footnote{
Strictly speaking, \eqref{sigmamodel} computes the index of the Dirac operator on the loop space
$\mathcal{L}X$ \cite{Witten:1986bf},\cite{Witten86}.
But when $X$ is hyperK\"ahler and consequently has the vanishing first Chern class, we can neglect the difference.}
twisted by a vector bundle $E_{w, Q}$ \cite{Li:2004ef}.
For any vector bundle $V$ over $X$, we can define the following formal sum:
\beq
\Lambda_t V = \oplus_{k \geq 0} t^k \Lambda^k V, \qquad
S_t V = \oplus_{k \geq 0} t^k S^k V,
\eeq
where $\Lambda^k$ and $S^k$ denote the $k$-th exterior and symmetric
product, respectively. Then the vector bundle $E_{w,Q}$ is defined as
\beq
E_{w,Q} := Q^{-\frac{d}{2}} \bigotimes_{n \geq 1}
\left( \Lambda_{-Q w^{n-1}} T_X^{*} \otimes \Lambda_{-Q w^n} T_X
\otimes S_{w^n} T_X^{*} \otimes S_{w^n} T_X \right),
\eeq
where $T_X$ is the holomorphic tangent bundle of $X$.
The index theorem tells that the elliptic genus is expressed in terms of the characteristic
classes:
\beq
Ell(X; w,Q) = \mathrm{index} \left( \bar\partial_{E_{w,Q}} \right) = \int_{X} \mathrm{ch} (E_{w,Q}) \mathrm{td}(X),
\eeq
where $\mathrm{ch}(V)$ denotes the Chern character of a vector bundle $V$,
and $\mathrm{td}(X)$ is the Todd class of $X$.
Let $\{ x_1, x_2, \cdots, x_d \}$ be the Chern roots of $T_X$.
Then we have
\beq
Ell(X; w,Q) = y^{-\frac{d}{2}} \int_{X} \prod_{n \geq 1} \prod_{k=1}^d
\frac{(1- Qw^{n-1} e^{-x_k})(1- Q^{-1} w^{n} e^{x_k})}
{(1 - w^n e^{-x_k})(1- w^n e^{x_k})} \cdot
\prod_{k=1}^d \frac{x_k}{1- e^{-x_k}},
\eeq
where the last factor comes from $\mathrm{td}(X)$.
When $X$ has a torus action with isolated fixed points $\{ p_1, p_2, \cdots p_\ell\}$,
we can employ the (Atiyah-Bott) localization theorem to evaluate
the integral over $X$ by the sum of the contribution from each fixed point \cite{Hollowood:2003cv}, \cite{Li:2004ef}.
Let $\{v_{i,1}, v_{i,2}, \cdots v_{i,d}\}$ be the weights of the torus action at $p_i$.
The localization theorem implies that
\beqa\label{localization}
Ell(X; w,Q) &=& Q^{-\frac{d}{2}} \sum_{i=1}^\ell
\frac{ \mathrm{ch} (E_{w,Q})[p_i] \cdot \mathrm{td}(X)[p_i] }{\prod_{k=1}^d v_{i,k}} \CR
&=& Q^{-\frac{d}{2}}\sum_{i=1}^\ell  \prod_{n \geq 1}\prod_{k=1}^d
\frac{(1- Qw^{n-1} e^{-v_{i,k}})(1- Q^{-1} w^{n} e^{v_{i,k})}}
{(1 - w^{n-1}  e^{-v_{i,k}})(1- w^n e^{v_{i,k}})} \CR
&=& Q^{-\frac{d}{2}} \sum_{i=1}^\ell \prod_{k=1}^d \frac{\theta_w(Q e^{-v_{i,k}})}{\theta_w(e^{-v_{i,k}})},
\eeqa
where $\mathrm{ch} (E_{w,Q})[p_i]$ and $\mathrm{td}(X)[p_i]$ denote the evaluation
at the fixed point $p_i$, and $\prod_{k=1}^d v_{i,k}$ is the contribution of the Euler class, which physically
comes from the Gaussian path integral of quantum fluctuation at the fixed point.

Now let us take the moduli space $\mathcal{M}_{n,k}$ of $U(n)$-instantons
with instanton number $k$ as the target space $X$.
A torus action on $\mathcal{M}_{n,k}$ is induced
from the action $(z_1,z_2) \to (t_1z_1,t_2 z_2)$ on $\mathbb{C}^2 \equiv \mathbb{R}^4$
and the Cartan torus of the gauge group $U(n)$.
We denote associated equivariant parameters $t_i$ ($t_1=t$, $t_2=q^{-1}$) and $u_\alpha~(\alpha=1,\cdots, n)$, respectively.
The isolated fixed points of the torus action on $\mathcal{M}_{n,k}$ are labelled by $n$-tuples
of Young diagrams $\vec\lam = \{ \lam^{(\alpha)} \}$ with $k = \vert \vec\lam \vert
= \sum_{\alpha=1}^n |\lam^{(\alpha)}|$. According to \cite{Nakajima}, \cite{Bruzzo:2002xf},
the equivariant character of the tangent space $T_{\vec\lam} \mathcal{M}_{n,k}$
at the fixed point $\vec\lam$ is given by
\beq\label{eqcha}
\chi (u_\alpha ; t_i)  = N^{*} K + t_1 t_2 K^{*} N - (1-t_1)(1-t_2) K^{*} K,
\eeq
where\footnote{
Our convention follows \cite{Nakajima}. To compare this one with some other in the literature, we have
to exchange $t_1$ and $t_2$, or make the transposition of the Young diagram.}
\beq
N := \sum_{\alpha=1}^n u_\alpha, \qquad K :=  \sum_{\alpha=1}^n u_\alpha \cdot
\left( \sum_{(i,j) \in \lam^{(\alpha)}} t_1^{1-j} t_2^{1-i} \right),
\eeq
and we denote the dual characters by $N^{*}$ and $K^{*}$.
We have the following combinatorial formula for $\chi (u_\alpha ; t_i) $ \cite{Nakajima}:
\beq\label{combi}
\chi (u_\alpha ; t_i)  =  \sum_{\alpha, \beta =1}^n \frac{u_\beta}{u_\alpha}
\left( \sum_{\square \in \lam^{(\alpha)}} t_2^{a_\alpha(\square)+1} t_1^{-\ell_\beta(\square)}
+ \sum_{\square \in \lam^{(\beta)}}  t_2^{-a_{\beta}(\square)} t_1^{\ell_\alpha(\square)+1} \right),
\eeq
In the original expression \eqref{eqcha}, there are both positive and negative contributions.
But after cancellation, only positive terms survive, and there are precisely $2nk$ terms,
which agrees with the (complex) dimensions of $\mathcal{M}_{n,k}$.
This means the tangent spaces at the fixed points are regular, and we can safely
read off the weights of the torus action from formula \eqref{combi}.
Combining \eqref{localization} and \eqref{combi}, we can compute the elliptic genus of
the moduli space $\mathcal{M}_{n,k}$ as follows:
\beq
Ell(\mathcal{M}_{n,k}; w,Q) = Q^{-nk} \sum_{\vert \vec{\lam} \vert =k }
 \prod_{\alpha,\beta =1}^n \left( \prod_{\square \in \lam^{(\alpha)}}
\frac{\theta_w(Q \frac{u_\beta}{u_\alpha} t_2^{a_\alpha(\square)+1} t_1^{-\ell_\beta(\square)})}
{\theta_w(\frac{u_\beta}{u_\alpha}t_2^{a_\alpha(\square)+1} t_1^{-\ell_\beta(\square)})}
\prod_{\square' \in \lam^{(\beta)}}
\frac{\theta_w(Q \frac{u_\beta}{u_\alpha} t_2^{-a_{\beta}(\square')} t_1^{\ell_\alpha(\square')+1} )}
{\theta_w(\frac{u_\beta}{u_\alpha} t_2^{-a_{\beta}(\square')} t_1^{\ell_\alpha(\square')+1} )}
\right).
\eeq
Thus the generating function of the elliptic genera of the instanton moduli spaces
\beqa
Z &:=& \sum_{k=0}^\infty p^{nk} Ell(\mathcal{M}_{n,k}; w,Q) \CR
&=& \sum_{\vec{\lam}} (p Q^{-1}) ^{n\vert \vec{\lam} \vert} \prod_{\alpha,\beta =1}^n
\mathcal{N}_{\lam^{(\alpha)},\lam^{(\beta)}} \left.\left( \frac{u_\beta}{u_\alpha}\right| t_2^{-1}, t_1^{-1}, w \right)
\eeqa
is expressed in terms of the elliptic lift of the Nekrasov factor $\mathcal{N}_{\lam^{(\alpha)},\lam^{(\beta)}}(u \vert q,s,w)$
defined by \eqref{ENekrasov}.
As was pointed out by E. Witten \cite{Witten:1986bf},\cite{Witten86},
the role of the Dirac operator in $K$ theory is replaced by the supersymmetric sigma model in elliptic cohomology.
Hence we may call the factor $\mathcal{N}_{\lam^{(\alpha)},\lam^{(\beta)}}(u \vert q,s,w)$ elliptic cohomological one.

To make similar computations for the affine Laumon space,
we use the equivalence of the instantons with parabolic structure (surface defect)
and the ramified instantons under the $\mathbb{Z}_n$ action on $\mathbb{C}^2$ by
\beq\label{orbifold}
(z_1, z_2) \to (z_1, \omega z_2)
\eeq
with $\omega^n=1$, \cite{MS},\cite{Biswas}. See also \cite{FFNR},\cite{Braverman:2010ef},\cite{FR}
for literatures directly related to the present case of full surface defect.
Computational aspects of the ramified instantons are described, for example,
in \cite{Kanno:2011fw},\cite{Nawata:2014nca},\cite{Bullimore:2014awa},\cite{Nekrasov:2017rqy}.
Note that this action is {\it different} from
\beq
(z_1, z_2) \to (\omega z_1, \omega^{-1} z_2),
\eeq
which defines a subgroup of $SU(2)$ and gives ALE space of $A_{n-1}$ type
as a resolution of $\mathbb{C}^2/\mathbb{Z}_n$.
According to \eqref{orbifold}, we replace $t_2 \to t_2^{\frac{1}{n}}$ and assign the charge $1/n$ to $t_2^{\frac{1}{n}}$.
We also put
\beq
u_\alpha = e^{a_\alpha} t_2^{-\frac{\alpha}{n}}
\eeq
so that $e^{a_\alpha}$ has the charge $\alpha/n$ to keep $u_\alpha$ neutral.
Let us decompose $N$ and $K$ by the $\mathbb{Z}_n$ orbifold charge.
For $N$ this is simply
\beq
N = \sum_{\alpha=1}^n t_2^{-\frac{\alpha}{n}} W_{\alpha}, \qquad W_{\alpha} = e^{a_\alpha}.
\eeq
On the other hand, we have
\beq
K = \sum_{\alpha=1}^n \sum_{\beta=1}^n \left( e^{a_\alpha} t_2^{\frac{1}{n}(1-\alpha-\beta)}
\sum_{m \geq 0} \sum_{i=1}^{\lam_{mn+\beta}^{(\alpha)}} t_1^{1-i} t_2^{-m} \right)
= \sum_{\gamma=1}^n t_2^{-\frac{\gamma}{n}} K_{\gamma}.
\eeq
Now the condition $1-\alpha - \beta \equiv - \gamma~\hbox{(mod $n$)}$ implies
$\alpha= \gamma - \beta +1$ for $1 \leq \beta \leq \gamma$
and $\alpha= \gamma - \beta + n + 1$ for $\gamma + 1 \leq \beta \leq n$.
Hence we find
\beq
K_{\gamma} = \sum_{\beta=1}^\gamma e^{a_{\gamma-\beta+1}} \sum_{m \geq 0}
\sum_{i=1}^{\lam_{mn+\beta}^{(\gamma - \beta +1)}} t_1^{1-i} t_2^{-m}
+  \sum_{\beta= \gamma+1}^n e^{a_{\gamma-\beta+1}} \sum_{m \geq 0}
\sum_{i=1}^{\lam_{mn+\beta}^{(\gamma - \beta +1)}} t_1^{1-i} t_2^{-m-1},
\eeq
where we have used $a_{\alpha + n} = a_{\alpha}$ and $\lam^{(\alpha+n)} = \lam^{(\alpha)}$.
Then $\mathbb{Z}_n$ invariant part of the \lq\lq orbifolded\rq\rq\ equivariant character \eqref{eqcha}
with $t_2 \to t_2^{\frac{1}{n}}$ is
\beq\label{Zninv}
\chi (u_\alpha ; q_i)^{\mathbb{Z}_n} = \sum_{\alpha=1}^n  \left (N_\alpha^{*} K_\alpha
+  t_1 K_{\alpha-1}^{*} N_\alpha - (1-t_1) K_\alpha^{*} K_\alpha + (1-t_1)  K_{\alpha-1}^{*} K_\alpha \right),
\eeq
with $K_0 \equiv K_n$.

The affine Laumon space $\mathcal{M}_{n, \vec{d}}\,$ gives a (small) resolution of the moduli space of instantons
with a full surface operator (or a maximal monodromy defect), which is described as follows.
Since the surface operator breaks $SU(n)$ gauge group\footnote{For the sake of simplicity, we omit the overall $U(1)$ symmetry.}
to $U(1)^{n-1}$, there are $(n-1)$ gauge fields on the surface $S \subset \mathbb{C}^2$.
We can define $(n-1)$ monopole numbers
\beq
m_i = \frac{1}{2\pi} \int_S F_i, \qquad (i=1, \cdots n-1)
\eeq
and also define $m_n := - m_1 - \cdots - m_{n-1}$. Then
the second Chern number of the ramified instantons is given by
\beq
c_2 = k + \sum_{i=1}^n \alpha_i m_i,
\eeq
where $\alpha_i$ measures the monodromy of the gauge fields around the codimension two defect\footnote{See also \cite{Nekrasov:2017gzb,CKL} for an alternative description in the Bethe/Gauge correspondence.}.
It is convenient to combine $k$, which we call instanton number, and the monopole numbers
to $\vec{d}=(d_1, d_2,\cdots,d_n)$ by
\beq\label{monopole}
d_0 = d_n = k, \quad d_{i} - d_{i-1} = m_i, \quad \sum_{i=1}^n m_i =0,
\eeq
so that the topological type of the affine Laumon space $\mathcal{M}_{n, \vec{d}}$ is labelled by $\vec{d}$.
Similarly to the ordinary instantons, the fixed points of the torus action on $\mathcal{M}_{n, \vec{k}}$
is labelled by $n$-tuples of partitions $\vec\lam$ which satisfy the condition
\beq
d_{i-1} = \sum_{\alpha \geq 1} \sum_{\alpha+ \beta \equiv i \atop
(\mathrm{mod}~n)} \lam_\alpha^{(\beta)}.
\eeq
It is known that the complex dimension of $\mathcal{M}_{n, \vec{k}}$ is $2n\sum_{i=1}^n d_i$.
We define the generating function of elliptic genera by
\beq
Z = \sum_{\vec{k}} \prod_{i=1}^n x_i^{2nm_i} Ell( \mathcal{M}_{n, \vec{k}} ; w,Q).
\eeq

According to \cite{FFNR},
the character formula for the affine Laumon space $\mathcal{M}_{n, \vec{k}}$ at the fixed point $\vec\lam$ is
\beqa
\mathrm{Ch} (\vec{a};  t_i)
&=&
( 1- t_1) \sum_{k=1}^n \sum_{1 \leq \ell} \sum_{1 \leq \tilde\ell}
e^{a_{k-\ell +1} - a_{k -\tilde\ell}} t_2^{\left( \floor{\frac{\tilde\ell -k}{n}} - \floor{\frac{\ell -k -1}{n}}\right)}
\sum_{i=1}^{\lam^{(k - \tilde\ell)}_{\tilde\ell}} t_1^{i-1} \sum_{j=1}^{\lam^{(k - \ell +1)}_\ell} t_1^{1-j} \CR
&& ~~ + t_1 \sum_{k=1}^n \sum_{ 1 \leq \tilde \ell } e^{a_k - a_{k -\tilde\ell}}
t_2^{\left(\floor{\frac{\tilde\ell-k}{n}} - \floor{-\frac{k}{n}}\right)}
\sum_{i=1}^{\lam^{(k-\tilde\ell)}_{\tilde\ell}} t_1^{i-1} \CR
&& - ( 1- t_1)  \sum_{k=1}^n  \sum_{ 1 \leq \ell } \sum_{1 \leq \tilde\ell}
e^{a_{k-\ell +1} - a_{k -\tilde\ell +1}}
t_2^{\left( \floor{\frac{\tilde\ell -k -1}{n}} - \floor{\frac{\ell -k -1}{n}}\right)}
\sum_{i=1}^{\lam^{(k - \tilde\ell +1)}_{\tilde\ell}} t_1^{i-1} \sum_{j=1}^{\lam^{(k - \ell +1)}_\ell} t_1^{1-j}  \CR
&& + \sum_{k=1}^n  \sum_{1 \leq \ell} e^{a_{k-\ell +1} - b_{k}}
t_2^{\left( \floor{\frac{-k}{n}} - \floor{\frac{\ell -k -1}{n}}\right)} \sum_{j=1}^{\lam^{(k - \ell +1)}_\ell} t_1^{1-j},
\eeqa
where we have substituted $\vec{b}=\vec{a}$ and $\vec{\mu} = \vec{\lam}$ in the original formula \cite[Prop. 4.15]{FFNR}.
Replacing $\ell \to  mn + \ell$ and $ \tilde\ell \to  \tilde m n + \tilde\ell$
with $0 \leq m,  \tilde m$ and $1 \leq \ell, \tilde\ell \leq n$, we can rewrite the character as follows:
\beqa
 \mathrm{Ch} (\vec{a};  t_i)
&=& (1- t_1) \sum_{k=1}^n  V_{k-1}^{*} (\vec{a}, \vec\lam) V_k (\vec{a}, \vec\lam)
+ t_1  \sum_{k=1}^n V_{k-1}^{*}  (\vec{a}, \vec\lam) W_k (\vec{a}) \CR
&&~~ - \sum_{k=1}^n   (1-t_1) V_k^{*} ( \vec{a}, \vec\lam) V_k (\vec{a}, \vec\lam)
+  \sum_{k=1}^n W_k^{*} (\vec{b}) V_k (\vec{a}, \vec\lam),
\label{Fformula}
\eeqa
where
$W_k (\vec{a}) := e^{a_k}$
and
\beq \label{BFNR}
V_k  (\vec{a}, \vec\lam) := \sum_{0 \leq m}  \sum_{\ell =1}^n \left(e^{a_{k-\ell +1}}
t_2^{-m -1 - \floor{\frac{\ell -k -1}{n}}} \sum_{j=1}^{\lam^{(k - \ell +1)}_{mn+ \ell}} t_1^{1-j} \right).
\eeq
To eliminate the floor function $\floor{\frac{\ell -k -1}{n}}$ in \eqref{BFNR}, we note that, for $1 \leq \ell \leq k$, it takes value $-1$, while,
for $k+1 \leq \ell \leq n$, it vanishes. Then we can see that \eqref{Fformula} exactly agrees with \eqref{Zninv}.

Thus the weights of the torus action on the tangent space of the affine Laumon space $\mathcal{M}_{n, \vec{d}}\,$
can be identified with $\mathbb{Z}_n$ invariant terms in \eqref{combi}, the generic term being $u_\beta/ u_\alpha \cdot t_1^{k} t_2^{\ell/n}$.
Taking the $\mathbb{Z}_n$ charges of $u_\alpha$, $t_1$ and $t_2^{1/n}$ into account, we can see
the selection rule for the $\mathbb{Z}_n$ invariant terms is $\beta - \alpha + \ell \equiv 0~(\mathrm{mod}~n)$,
which is nothing but the selection rule in \eqref{EGfactor}. The monomial $x$-factor in \eqref{EG} also matches
the relation of $\vec{d}$ and the monopole numbers \eqref{monopole}.

\section{Conjecture and various limits
\label{claim}}

We are now ready to make the main claim of the present paper, which is supported by the consideration of the previous sections.

\paragraph{Conjecture.}\framebox{\parbox{12cm}
{The ELS-function satisfies non-stationary equations with (yet not completely understood) quantum Dell Hamiltonians.}}

\vspace{1cm}

Equivalently, we expect that this is the instanton partition function with full monodromy defect in the $6d$ SYM theory with adjoint matter hypermultiplet.

By the AGT correspondence, it is the 2-point conformal block on torus with one degenerate field in a double elliptic version of the quantum toroidal algebra. Let us see how the associated parameters look like in these pictures. Note that the limits and symmetries below require some proper normalization of the ELS-functions as we explained in sec.\ref{u1}.

The ELS-function $\mathfrak{P}_n^{E.G.} { (x_i ; p \vert y_i ; s \vert q,t,w)}$ depends on the five parameters $(p,s,q,t,w)$:
\begin{itemize}
\item[$q$, $s$] are two $\Omega$-background deformation parameters on the gauge theory side, $s$ governs the non-stationarity. The limit of $s\to 1$ reduces the system to the quantum integrable system, and the non-stationary Dell equation to the eigenvalue Dell Hamiltonian problem. On the algebra side, they rescale the dimensions of the operators.
\item[$t$] is the central charge parameter on the algebra side, and the coupling constant parameter on the integrable side. The integrable system becomes free upon $t\to 1$. On the gauge theory side, it is related to the mass of the adjoint hypermultiplet, $t=e^{-m}$.
\item[$p$] is the elliptic parameter that controls the coupling in the gauge theory (the bare torus and the bare charge). On the integrable side, it is associated with the torus where the coordinates live. On the algebra side, it is associated with the torus where the $2d$ fields in the $4d$ limit live. In the limit $p\to 0$, the instanton corrections disappear, and one gets the perturbative limit.
\item[$w$] is the elliptic parameter that is associated with the Kaluza-Klein torus in the gauge theory (remind that one considers Seiberg-Witten $6d$ theory with two dimensions compactified onto a $2d$ torus). On the integrable side, it is associated with the torus where the momenta live.
\end{itemize}

Thus the correspondence table between parameters of ELS- and
$6d$ Nekrasov functions looks as follows:
\be
w=e^{2\pi i\hat\tau},\quad p=e^{2\pi i\tau},\quad t=e^{-m},\quad q=e^{-2\pi i\epsilon_1},\quad s=e^{-2\pi i\epsilon_2}
\ee
To this, we add an expectation of a peculiar pair of trialities: the triality of $(q,s,t)$ and the elliptic parameter triality of $(p/t,w/t,t/(qs))$. Note that this parametrization is in terms of dimensionless parameters. Actually, they depend on a length scale $L$: $\epsilon_{1,2}\to L\epsilon_{1,2}$, $m\to Lm$. Taking into account the behaviour of these parameters under the modular transformation $\hat\tau\to -1/\hat\tau$ (see\footnote{There is a misprint in \cite[Eq.(6.38)]{AKMMSZ}: $P$ contains an additional factor $e^{\pi i(\epsilon_1+\epsilon_2)}$, which, however, does not influence the calculations.} \cite[Eqs.(6.36)-(6.38)]{AKMMSZ}) and that $\hat\tau$ is the ratio of radii of the 5th and 6th dimensions under compactification, it is natural to choose this length to be one of these radii.

\bigskip

This association implies the following degenerations:

\begin{itemize}
\item[$w\to 0$.] It drives the system to the $5d$ gauge theory with adjoint matter. In this limit, we get the Shiraishi function, which is the instanton partition function with a monodromy defect in the $5d$ theory with adjoint matter hypermultiplet.

By the AGT correspondence, it is the 2-point conformal block on torus with one degenerate field in $q$-Virasoro algebra.

On the integrable side, this function has to satisfy a non-stationary elliptic Ruijsenaars equation, in accordance with the general rule \cite{MMMsurfop}. This is, indeed, the case, see \cite{Shi}, where various further limits are also studied.

The $s\to 1$ limit of the Shiraishi function describes the eigenfunctions of the elliptic Ruijsenaars Hamiltonians, as it should be, since this the Nekrasov-Shatashvili limit.
\item[$s\to 1$.] In this limit, the ELS-function is the eigenfunction of the Dell Hamiltonians.
\item[$p\to 0$.] It is the perturbative limit, which is described by the dual elliptic Ruijsenaars system. The corresponding ELS-functions are given by (\ref{p0limit}), $Z (x_i \vert y_i  \vert q,t,w)$.
\end{itemize}

It is unclear if the Hamiltonians coincide with the recent suggestion in \cite{Koroteev:2019gqi}: in the last limit and $n=2$, their eigenfunctions
 \cite[eqs.(67)-(68)]{AKMMdell1} look rather different from (\ref{p0limit}).

\section{Conclusion
\label{conc}}

To conclude, we made one more step towards a complete solution of the Dell system.
According to the general approach of network models, they are obtained by
closing (compactifying) the DIM network in the both directions,
horizontal and vertical.
Making this directly for representation theory of the DIM algebra is still to be done,
we used a somewhat different approach from \cite{AKMMdell1,AKMMdell3}
based on the Shiraishi functor,
which can naturally provide eigenfunctions of the still-not-fully known Hamiltonians.
This approach is analogous to consideration of hypergeometric series {\it per se}
without explicit reference to their origins/interpretation in terms of
representation theory of $SL(2,\mathbb{R})$ \cite{GGPS}, and, in many respects, it is considerably simpler.
As we explained, for describing the Dell systems, one actually needs an elliptic lift
of the ordinary Shiraishi series, where the Pochhammer symbols are changed for their
elliptic counterparts. We discussed it in some detail, paying a special attention
at various limits and particular cases.
In fact, the elliptic substitute of the Pochhammer symbol is only a very particular deformation:
one can actually change it for an arbitrary function, with many properties of
the Shiraishi functions still preserved \cite{AKMMdell3}. However, this is the elliptic deformation that very non-trivially
preserves the relation to symmetric functions (see sec.3).
In this paper, we concentrate on the elliptic (theta-function) deformation by this reason and
since {\it it} is supposed to lead to the theory of Dell systems and a related
description of $6d$ SUSY gauge theory with adjoint matter hypermultiplet compactified on the two-dimensional torus.

\section*{Acknowledgements}

We appreciate useful discussions with M.Fukuda, Y.Ohkubo, J.Shiraishi and Y.Zenkevich.

Our work is supported in part by Grants-in-Aid for Scientific Research
(17K05275) (H.A.), (15H05738, 18K03274) (H.K.) and JSPS Bilateral Joint Projects (JSPS-RFBR collaboration)
``Elliptic algebras, vertex operators and link invariants'' from MEXT, Japan. It is also partly supported by the grant of the Foundation for the Advancement of Theoretical Physics ``BASIS" (A.Mir., A.Mor.), by  RFBR grants 19-01-00680 (A.Mir.) and 19-02-00815 (A.Mor.), by joint grants 19-51-53014-GFEN-a (A.Mir., A.Mor.), 19-51-50008-YaF-a (A.Mir.), 18-51-05015-Arm-a (A.Mir., A.Mor.), 18-51-45010-IND-a (A.Mir., A.Mor.). The work was also partly funded by RFBR and NSFB according to the research project 19-51-18006 (A.Mir., A.Mor.).
We also acknowledge the hospitality of KITP and partial support by the National Science Foundation under Grant No. NSF PHY-1748958.

\end{document}